\documentclass[10pt]{iopart}
\usepackage{iopams,graphicx}

\begin{document}

\title{Fractional Brownian motion in presence of two fixed adsorbing
boundaries}

\author{
G Oshanin}

\address{ Laboratoire de Physique Th{\'e}orique de la
Mati{\`e}re Condens{\'e}e, Universit{\'e} Paris 6, 4 Place Jussieu,
75252 Paris, France}

\eads{ \mailto{oshanin@lptmc.jussieu.fr}}

\begin{abstract}
We study the long-time asymptotics of the probability $P_t$ 
that the  Riemann-Liouville
fractional Brownian motion with Hurst index $H$ 
does not escape from a fixed interval $[-L,L]$ up to time $t$. We show that for any $H \in ]0,1]$, 
for both subdiffusion and superdiffusion regimes, this probability obeys $\ln(P_t) \sim - t^{2 H}/L^2$, i.e. may decay  slower than exponential (subdiffusion) or faster than exponential (superdiffusion). 
 This implies that survival probability $S_t$ of particles undergoing fractional Brownian motion in a one-dimensional system with randomly placed traps follows $\ln(S_t) \sim - n^{2/3} t^{2H/3}$ as $t \to \infty$, where $n$ is the mean density of traps.
\end{abstract}



\section{Introduction}

Consider stochastic process $X_t^H$ defined as
\vspace{0.1 in}
\begin{equation} 
\label{def} \fl X_t^H = \frac{1}{\Gamma(H + 1/2)} \int^t_0
\frac{d\tau \zeta(\tau)}{(t - \tau)^{1/2 - H}},
\end{equation} \vspace{0.1 in}
where $\zeta(\tau)$ is Gaussian, delta-correlated noise and
$H$  - the Hurst index, is a real number in $]0,1]$.

The process in Eq.(\ref{def}) is called the Riemann-Liouville
fractional Brownian motion (fBm), since $X_t^H$ is the solution of
the Langevin equation in which random force is the fractional
Riemann-Liouville derivative of Gaussian noise. The fBm
was first formulated in Kolmogorov's paper \cite{k} and later in the
papers of L\'evy \cite{l}, but first systematic analysis belongs to
Mandelbrot and Van Ness \cite{m}.

The fBm $X_t^H$ is a continuous-time Gaussian process starting at
zero, with mean zero, having the correlation function:
\vspace{0.1 in}
\begin{equation} 
\label{e} \fl E[X_t^H X_s^H] = \frac{(H+1/2) t^{H-1/2}
s^{H+1/2}}{\Gamma^2(H+3/2) } \,_2F_1\left(\frac{1}{2}-H, 1;
H+\frac{3}{2}; \frac{s}{t}\right),
\end{equation} \vspace{0.1 in}
where $s < t$ and $F$ denotes the Gauss hypergeometric function, and the
variance:
\vspace{0.1 in}
\begin{equation}
\label{v} \fl E\left[\left(X_t^H\right)^2\right] = \frac{t^{2H}}{2H
\Gamma^2(H+1/2)}.
\end{equation} \vspace{0.1 in}
Equations (\ref{e}) and (\ref{v}) signify that $X_t^H$ has
self-similar but not independent increments: for $H > 1 / 2$ the
increments of the process are positively correlated and the fBm
shows a superdiffusive behavior, while for $H < 1/2$ the increments
of the process are negatively correlated and one deals with
subdiffusion. In case $H = 1/2$ one recovers standard Brownian
motion with diffusion coefficient $D = 1/2$.

Here we study a simple first-passage problem for the fBm, whose
general understanding is a basic aspect of stochastic processes
\cite{katja,sid}. Namely, we analyze the asymptotic long-time
behavior of the probability $P_t$ that the fBm $X_t^H$ in
Eq.(\ref{def}) does not escape from the interval $[-L,L]$ up to time $t$. We note
that contrary to a single boundary case, for which several
rigorous results are available \cite{sin,mol}, understanding of the
two boundaries case is still rather controversial:\\
(a) One source of confusion stems from a recent tendency of
describing all wealth of naturally occurring anomalous diffusive
processes in terms of the so-called
 \textit{fractional} diffusion equation, fractional Fokker-Planck
and other fractional differential equations, regardless of the
origin, intrinsic correlations and physics underlying these
processes. If non-Markovian fBm-type processes are indeed described
by fractional diffusion equations \cite{coffey}, one expects that
$P_t$ will have an \textit{algebraic} tail. This would imply that
the distribution of the adsorption or first exit time from a fixed
interval will not have all moments. This is, of course, a rather
 counterintuitive conclusion.\\
(b) Survival of a tagged bead of an infinitely long Rouse polymer
chain, (whose dynamics is an fBm-type process with $H = 1/4$), in
presence of two adsorbing boundaries has been discussed in
Ref.\cite{nech}. Using a path-integral formulation with an exact measure
of trajectories of such a bead \cite{b}, and adapting a classic
method of images, it was shown that the bead's survival probability
obeys
\vspace{0.1 in}
\begin{equation}
\label{b} \fl - \ln\left(P_t\right) \sim \frac{t^{1/2}}{L^2},
\end{equation} \vspace{0.1 in}
i.e., is described by a stretched-exponential function of time.\\
(c) Numerical simulations of a tagged particle
dynamics in a one-dimensional hard-core lattice gas  - another
fBm-type process with $H = 1/4$, - in presence of two adsorbing
boundaries \cite{bunde}, and more recent simulations of dynamics of
a tagged bead of a finite Rouse chain between two traps
\cite{kardar}, suggested both a faster decay of the survival
probability:
\vspace{0.1 in}
\begin{equation}
\label{c} \fl - \ln\left(P_t\right) \sim \frac{t}{L^4}.
\end{equation} \vspace{0.1 in}
We set out to show here that, for $0 < H \leq 1$, i.e., both in the
subdiffusive and superdiffusive regimes, the survival probability
$P_t$ obeys
\vspace{0.1 in}
\begin{equation}
\label{main} \fl - \ln\left(P_t\right) \sim {\rm const} \frac{t^{2
H}}{L^2},
\end{equation} \vspace{0.1 in}
which expression can be rewritten, taking advantage of Eq.(\ref{v}), as
\vspace{0.1 in}
\begin{equation}
\label{main1} \fl - \ln\left(P_t\right) \sim {\rm const}
\frac{E\left[\left(X_t^H\right)^2\right]}{L^2}.
\end{equation} \vspace{0.1 in}
Note that our result in Eq.(\ref{main}) confirms
Eq.(\ref{b}) and contradicts (a) and (c), Eq.(\ref{c}).

\section{Basic equations}

Since we are concerned with the large-$t$ behavior, it will not matter
much how we define $\zeta(\tau)$ - as a continuous in
time function or as a discrete process, provided that we keep all
essential features of noise. We thus divide, at a fixed $t$, the
interval $[0,t]$ into $N$ ($N \gg 1$) small subintervals $\Delta$,
(such that $\Delta N \equiv t$), and assume that within the $k$-th
subinterval, $k=0,1, \ldots, N-1$, the noise $\zeta(\tau)$
is constant and equal to $\zeta_k/\sqrt{\Delta}$, where
$\{\zeta_k\}$ are independent random variables with normal
distribution $N[0,1]$.

Then, the Riemann-Liouville fBm in Eq.(\ref{def}) can be written down as a "weighted"
sum of independent random variables:
\vspace{0.1 in}
\begin{eqnarray}
\fl X_t^H = \frac{1}{\Gamma(H + 1/2)} \sum_{k=0}^{N-1}
\frac{\zeta_k}{\sqrt{\Delta}} \int^{(k+1) \Delta}_{k \Delta} \frac{d
\tau}{(t - \tau)^{1/2 - H}} =  \sum_{l=1}^{N} \sigma_l \, \zeta_{N -
l},
\end{eqnarray} \vspace{0.1 in}
where  $\sigma_l$ is a non-random function:
\vspace{0.1 in}
\begin{equation}
\label{sigma} \fl \sigma_l = \frac{\Delta^{H}}{\Gamma(H + 3/2)}
\left[l^{H+1/2} - (l - 1)^{H+1/2}\right].
\end{equation} \vspace{0.1 in}
Note that $\sigma_l$ is a \textit{monotonically decreasing} function
of $l$ for $H < 1/2$, a constant for $H = 1/2$, and a
\textit{monotonically increasing} function of $l$ for $H > 1/2$.

Now, let ${\rm rect}_L(X)$ denote the rectangular function:
\vspace{0.1 in}
\begin{equation} \fl
{\rm rect}_L(X) = \cases{
1, & ~~ $|X| < L$,\\
1/2, & ~~ $X=\pm L$, \\
0,  & ~~ $|X| > L$\\} \label{indicator}
\end{equation} \vspace{0.1 in}
Representing ${\rm rect}_L(X)$ via its Fourier transform:
\vspace{0.1 in}
\begin{equation}
\fl {\rm rect}_L(x) =  \int_{-\infty}^{\infty} \frac{d k}{\pi}
\frac{\sin(L k)}{k} \exp\left[i k X\right],
\end{equation} \vspace{0.1 in}
we may now write down the indicator function ${\rm I}\Big({\rm
max}|X_t^H| \leq L\Big)$ of the event that an $N$-step trajectory
 $X_t^H$ did not leave the interval $[-L,L]$ as the following $N$-fold integral:
\vspace{0.1 in}
\begin{equation}
\fl {\rm I}\Big({\rm max}|X_t^H| \leq L\Big) =  
\int_{-\infty}^{\infty} \ldots \int_{-\infty}^{\infty} \prod_{l=1}^N
\frac{dk_l}{\pi} \; \frac{\sin(L k_l )}{k_l} \; \exp\left[i \; \sigma_l \; \zeta_{N-l}
\; \sum_{j=1}^{N - l + 1} k_j\right].
\end{equation} \vspace{0.1 in}
Averaging the latter equation, we have then that the probability
$P_N$ of this event is given by
\vspace{0.1 in}
\begin{equation}
\fl P_N =  \int_{-\infty}^{\infty} \ldots
\int_{-\infty}^{\infty} \prod_{l=1}^N \frac{d k_l}{\pi} \;  \frac{\sin(L k_l )}{k_l} \;
\exp\left[- \frac{\sigma_l^2}{2} \; \left(\sum_{j=1}^{N - l + 1}
k_j\right)^2\right].
\end{equation} \vspace{0.1 in}
Next, changing next the integration variables:
\begin{eqnarray}
Y_1 &=& k_1 + k_2 + \ldots + k_N, \nonumber\\
Y_2 &=& k_1 + k_2 + \ldots + k_{N-1}, \nonumber\\
Y_3 &=& k_1 + k_2 + \ldots + k_{N-2}, \nonumber\\
\ldots \nonumber\\
Y_N &=& k_1,
\end{eqnarray}
we obtain
\vspace{0.1 in}
\begin{equation}
\label{u} \fl P_N =  \int_{-\infty}^{\infty} \ldots
\int_{-\infty}^{\infty} \prod_{l=1}^{N} \frac{dY_l}{\pi} \; \frac{\sin\left(L
(Y_l - Y_{l+1}) \right)}{Y_l - Y_{l+1}} \; \exp\left[- \frac{1}{2} \sum_{l=1}^N \sigma_l^2
Y_l^2\right], \; Y_{N+1} \equiv 0.
\end{equation} \vspace{0.1 in}
Now, it is expedient to use the following integral identity for the sinc-function:
\vspace{0.1 in}
\begin{equation}
\label{uu} \fl \frac{\sin\left(L \left(Y_l - Y_{l+1}\right)
\right)}{Y_l - Y_{l+1}} = \frac{1}{2} \int_{-L}^L dX \exp\left[i X
\left(Y_l - Y_{l+1}\right)\right].
\end{equation} \vspace{0.1 in}
Plugging Eq.(\ref{uu}) into Eq.(\ref{u}), and performing
integrations over $\{Y_l\}$, we finally arrive at the following meaningfull representation of the survival probability:
\vspace{0.1 in}
\begin{eqnarray}
\label{f} \fl P_N = \int_{-L}^{L} \ldots \int_{-L}^{L}
\prod_{l=1}^{N} \frac{dX_l}{\sqrt{2 \pi} \sigma_l} \; \exp\left[- \sum_{l=1}^N \frac{\left(X_l -
X_{l-1}\right)^2}{2 \sigma_l^2} \right] = \nonumber\\
= \int_{-L}^{L} \ldots \int_{-L}^{L} \prod_{l=1}^{N} \; dX_l
\exp\left[- \sum_{l=1}^N \left(X_l - X_{l-1}\right)^2/2 \sigma_l^2
\right] / \nonumber\\
/ \int_{-\infty}^{\infty} \ldots \int_{-\infty}^{\infty}
\prod_{l=1}^{N} dX_l \; \exp\left[- \sum_{l=1}^N \left(X_l -
X_{l-1}\right)^2/2 \sigma_l^2 \right], \; X_0 \equiv 0.
\end{eqnarray}

Note that the integrand in Eq.(\ref{f}) is an 
analog of the Wiener measure for the Riemann-Liouville fractional
Brownian motion.

\section{Hurst index $1/4 < H \leq 1/2$: fBm as a Brownian motion in an expanding cage.}

Change the integration variables $X_l = \sigma_l x_l$. Then,
Eq.(\ref{f}) reads
\vspace{0.1 in}
\begin{equation}
\label{ff} \fl P_N = 
\int_{-L/\sigma_1}^{L/\sigma_1} \int_{-L/\sigma_2}^{L/\sigma_2}
\ldots
\int_{-L/\sigma_N}^{L/\sigma_N} \prod_{l=1}^{N} \frac{dx_l}{\sqrt{2 \pi}} \, \exp\left[- \frac{1}{2}
\sum_{l=1}^N \left(x_l - \frac{\sigma_{l -1}}{\sigma_l}
x_{l-1}\right)^2 \right], 
\end{equation} \vspace{0.1 in}
where $x_0 \equiv 0$.
Next, we represent
\vspace{0.1 in}
\begin{equation}
\fl \sum_{l=1}^N \left(x_l - \frac{\sigma_{l -1}}{\sigma_l}
x_{l-1}\right)^2 = \sum_{l=1}^N \left(x_l - x_{l-1}\right)^2 +
F\left(\{x_l\}\right),
\end{equation} \vspace{0.1 in}
where
\vspace{0.1 in}
\begin{equation}
\label{kk} \fl F\left(\{x_l\}\right) = 2 \sum_{l = 1}^N \left(1 -
\frac{\sigma_{l-1}}{\sigma_l}\right) x_{l-1} \left(x_l -
x_{l-1}\right) + \sum_{l = 1}^N \left(1 -
\frac{\sigma_{l-1}}{\sigma_l}\right)^2 x_{l-1}^2.
\end{equation}
Now, note that it follows from Eq.(\ref{sigma}) that $\sigma_l \sim l^{H-1/2}$ and $(1 -
\sigma_{l-1}/\sigma_l) \sim 1/l$ as $l \to \infty$. Hence, for any
$H
> 0$ and any $N$, the second sum on the right-hand-side of Eq.(\ref{kk}) is
bounded:
\vspace{0.1 in}
\begin{eqnarray}
\fl {\rm max}\left|  \sum_{l = 1}^N \left(1 -
\frac{\sigma_{l-1}}{\sigma_l}\right)^2 x_{l-1}^2\right| \leq  L^2 C, \;\;\; C \equiv 
\sum_{l=1}^{\infty} \sigma_{l-1}^{-2} \left(1 -
\frac{\sigma_{l-1}}{\sigma_l}\right)^2.
\end{eqnarray}
On the other hand, an elementary analysis shows that for $0 < H <
1/2$ the maximal absolute value of the first sum on the right-hand-side
of Eq.(\ref{kk}) grows with $N$:
\vspace{0.1 in}
\begin{eqnarray}
\fl {\rm max}\left|\sum_{l = 1}^N \left(1 -
\frac{\sigma_{l-1}}{\sigma_l}\right) x_{l-1} \left(x_l -
x_{l-1}\right)\right|  \sim
L^2 C_1 N^{1 - 2 H},
\end{eqnarray} \vspace{0.1 in}
where $C_1$ is $N$-independent constant.

Consequently, the survival probability $P_N$ obeys the following double-sided inequality:
 \vspace{0.2 in}
\begin{equation}
\label{ineqq}
\fl \exp\left[- L^2 \left(C_1 N^{1 - 2 H} + C\right)\right] \,
\Psi_N \, \leq P_N \, \leq \, \exp\left[L^2 \left(C_1 N^{1 - 2 H} +
C\right)\right] \, \Psi_N,
\end{equation} \vspace{0.1 in}
in which $\Psi_N$ is given explicitly by
\vspace{0.1 in}
\begin{equation}
\label{fff} \fl \Psi_N = 
\int_{-L/\sigma_1}^{L/\sigma_1} \int_{-L/\sigma_2}^{L/\sigma_2}
\int_{-L/\sigma_3}^{L/\sigma_3}\ldots
\int_{-L/\sigma_N}^{L/\sigma_N} \prod_{l=1}^{N} \frac{dx_l}{\sqrt{2 \pi}} \exp\left[-
\sum_{l=1}^N \frac{\left(x_l - x_{l-1}\right)^2}{2} \right], 
\end{equation} \vspace{0.1 in} 
with $x_0 \equiv 0$.

One notices now that Eq.(\ref{fff}) describes the probability that
an $N$-step Brownian motion trajectory $x_l$, starting at the origin, 
does not escape from
the interval whose boundaries move deterministically \textit{away
from the origin} as $\pm L/\sigma_l$, $l = 1,2, \ldots, N$. This classical problem has
been extensively studied in the probability theory (see Ref.\cite{nov}
and references therein). A lucid derivation of main results and
description of different approaches can be also found in
Ref.\cite{kr}.

At sufficiently large $N$, $\Psi_N$ obeys \cite{nov} (note that
we appropriately change the notations):
\vspace{0.1 in}
\begin{eqnarray}
\label{lll} \fl \Psi_N \sim \exp\left[- \frac{\pi^2}{8 L^2}
\sum_{l=1}^N \sigma_l^2 \right] &\sim& \exp\left[- \frac{\pi^2}{16 H
\Gamma^2(H+1/2)} \frac{\left(\Delta N\right)^{2 H}}{L^2} \right] = \nonumber\\ &=& \exp\left[- \frac{\pi^2}{8} \frac{E\left[\left(X_t^H\right)^2\right]}{L^2}
 \right]
\end{eqnarray} \vspace{0.1 in}
Note now that for $1/4 < H < 1/2$, $\Psi_N$ in Eq.(\ref{lll}) 
decays faster than $\exp[- N^{1 - 2H}]$, and hence, in virtue of the inequality in Eq.(\ref{ineqq}), 
$\Psi_N$ 
determines the decay
of the survival probability $P_N$, which yields the result in
Eq.(\ref{main}).

\section{Hurst index $1/2 < H \leq 1$: fBm as a Brownian motion in a shrinking cage.}

Consider next the superdiffusive case when $1/2 < H < 1$. One
notices that here $\sigma_l \sim l^{H-1/2} \to \infty$ as $l
\to \infty$, while $(1 - \sigma_{l-1}/\sigma_l) \sim 1/l$, and
hence, $F\left(\{x_l\}\right)$, Eq.(\ref{kk}), is bounded
by a constant for any N, i.e.,
\vspace{0.1 in}
\begin{eqnarray}
\fl {\rm max}|F\left(\{x_l\}\right)| \leq 2 L^2 C_2, \;\;\; C_2 &\equiv& 
\sum_{l = 1}^{\infty} \left(1 - \frac{\sigma_{l-1}}{\sigma_l}\right)
\sigma_{l-1}^{-1} \left(\frac{1}{\sigma_l} +
\frac{1}{\sigma_{l-1}}\right) + \nonumber\\ \fl &+& \frac{1}{2} \sum_{l =
1}^{\infty} \left(1 - \frac{\sigma_{l-1}}{\sigma_l}\right)^2
\sigma_{l-1}^2,
\end{eqnarray} \vspace{0.1 in}
Consequently, for $1/2 < H  \leq 1$, the survival probability $P_N$
obeys:
\vspace{0.1 in}
\begin{equation} 
\fl \exp\left[- L^2 C_2\right] \, \Psi_N \, \leq P_N \, \leq \,
\exp\left[L^2 C_2 \right] \, \Psi_N,
\end{equation} \vspace{0.1 in}
where $\Psi_N$ is defined by Eq.(\ref{fff}).

Contrary to the situation discussed in the previous subsection, here, i.e.
for $1/2 < H \leq 1$, $\Psi_N$ describes the probability that an
$N$-step Brownian motion trajectory $x_l$, commencing at the origin,
 does not escape from the
interval whose boundaries move deterministically \textit{towards}
the origin, i.e. that it survives in a \textit{shrinking} cage.

In this subsection we estimate the long-time asymptotical behavior of
$\Psi_N$ using an adiabatic approximation described in Ref.\cite{kr}. To
ascertain the accuracy of this approach, in the next section we will
present the results of a more rigorous analysis.

To define an asymptotic behavior of $\Psi_N$, consider the solution
of a diffusion equation
\vspace{0.1 in}
\begin{equation}
\label{aa} \fl \frac{\partial P(X,t)}{\partial t} = \frac{1}{2
\Delta} \frac{\partial^2 P(X,t)}{\partial X^2}, \;\;\; P(X,t=0) =
\delta(X), \vspace{0.1 in}
\end{equation}
subject to the boundary conditions
\vspace{0.1 in}
\begin{equation}
\fl P(X = \pm L(t),t) = 0,
\end{equation} \vspace{0.1 in}
where $L(t) = L/\sigma_t$ and $\sigma_t \sim  \sqrt{\Delta} \; t^{H
- 1/2}/\Gamma(H+1/2)$, Eq.(\ref{sigma}).

The basic idea behind the adiabatic approximation is that, if the
cage expands or shrinks sufficiently slowly, the density
distribution approaches the same form as in the fixed-cage case,
except that the parameters in this probability distribution acquire
time dependence to satisfy the moving boundary condition \cite{kr}.
Within this approximation, one takes
\vspace{0.1 in}
\begin{equation}
\label{a} \fl P(X,t) \sim f(t) \cos\left(\frac{\pi X}{2 L(t)
}\right) = f(t) \cos\left(\frac{\pi X}{2 L} \frac{ \sqrt{\Delta}
}{\Gamma(H+1/2)} t^{H-1/2}\right),
\end{equation} \vspace{0.1 in}
where the amplitude $f(t)$ is to be determined. Substituting
Eq.(\ref{a}) into Eq.(\ref{aa}), one finds
\vspace{0.1 in}
\begin{equation}
\label{aaa} \fl \frac{d f(t)}{d t} = \left(\frac{\pi^2}{8 \Delta
L^2(t)}\right) f(t) - \left(\frac{\pi X}{2 L^2(t)}\right)
\tanh\left(\frac{\pi X}{2 L(t)}\right) \frac{d L(t)}{d t} f(t).
\end{equation} \vspace{0.1 in}
Noticing next that the second term on the right-hand-site of
Eq.(\ref{aaa}) decays faster than the first one, we find
\vspace{0.1 in}
\begin{equation}
\fl f(t) \sim \exp\left[- \frac{\pi^2}{8 \Delta} \int_0^t \frac{d
\tau}{L^2(\tau)} \right] \sim \exp\left[- \frac{\pi^2}{16 H
\Gamma^2(H+1/2)} \frac{t^{2 H}}{L^2} \right],
\end{equation} \vspace{0.1 in}
and hence, $\Psi_t$ follows
\begin{eqnarray}
\label{psi} \fl \Psi_t \sim \int^{L(t)}_{-L(t)} dX P(X,t) &=&
\frac{4}{\pi} L(t) \exp\left[- \frac{\pi^2}{16 H \Gamma^2(H+1/2)}
\frac{t^{2 H}}{L^2} \right] \sim \nonumber\\
&\sim& \exp\left[- \frac{\pi^2}{8} \frac{E\left[\left(X_t^H\right)^2\right]}{L^2}
 \right]
\end{eqnarray} 

As in the case $1/4 < H < 1/2$, $\Psi_t$ determines the decay
of the survival probability $P_t$ and hence, we obtain the result in
Eq.(\ref{main}). Note also that the exponential term in
Eq.(\ref{psi}), describing leading behavior in the superdiffusive regime with $1/2 < H \leq
1$, is \textit{exactly} the same as the one in Eq.(\ref{lll}),
describing subdiffusion, but here it defines a \textit{faster than
exponential} decay of the survival probability.

\section{Arbitrary Hurst index, $0 < H \leq 1$: Upper and lower bounds.}

In this section, our analysis will be based on the following two keystones:\\
(i) The probability $P\left({\rm max|X_t^{H=1/2}| \leq M}\right)$
that Brownian motion $X_t^{H=1/2}$, starting at the origin, does not
leave an interval $[-M,M]$ up to time $t$, $t$ sufficiently large, is given by (see, e.g.,
Ref.\cite{feller}):
\vspace{0.1 in}
\begin{equation}
\label{w} \fl P\left({\rm max|X_t^{H=1/2}| \leq M}\right) \sim
\exp\left[- \frac{\pi^2}{8 M^2} t\right].
\end{equation}
\vspace{0.1 in}
(ii) A \textit{fundamental} property of $P_N$, Eq.(\ref{f}): 
$P_N = P_N(\sigma_1,\sigma_2, \ldots,\sigma_N)$ is a
\textit{monotonically decreasing} function of any variable $\sigma_k$, i.e.
\vspace{0.1 in}
\begin{equation}
\label{in}
\fl \frac{\partial P_N}{\partial \sigma_k} \leq 0, \;\;\; 0 \leq
\sigma_k < \infty \vspace{0.2 in}
\end{equation} \vspace{0.1 in}
Equation (\ref{in}) signifies that replacing any or all $\sigma_k$ by $\Sigma(k)$, such that
$\sigma_k \leq \Sigma(k)$ for any $k$, we will decrease the survival
probability and arrive at the \textit{lower} bound on $P_N$; if, on
contrary, we will replace one or all $\sigma_k$ by
$\tilde{\Sigma}(k)$, such that $\sigma_k \geq \tilde{\Sigma}(k)$ for
any $k$, we will \textit{increase} the survival probability and
obtain an \textit{upper} bound on $P_N$.

We are unaware of any statement similar to (ii) made in the
literature, apart of a less general "Pascal Principle" \cite{mor}.
It might be thus instructive to demonstrate first its validity.

Let us first single out terms dependent on $X_k$ and $X_{k-1}$ in Eq.(\ref{f}) writting down $P_N$ formally as
\vspace{0.1 in}
\begin{eqnarray}
\label{fi} \fl P_N = 
 &&\int_{-L}^{L}
\ldots \int_{-L}^{L} \prod_{l=1, l\neq k-1, k}^{N} \frac{dX_l}{\sqrt{2 \pi} \sigma_l} \,  B\left(\{X_{l}\}\right) \times \nonumber\\
\fl &\times& \left(\int_{-L}^{L} \frac{dX_k}{\sqrt{2 \pi} \sigma_k} \int_{-L}^{L}
\frac{dX_{k-1}}{\sqrt{2 \pi} \sigma_{k-1}} \,  B'\left(X_{k-1},X_k,X_{k+1}\right)\right), 
\end{eqnarray} \vspace{0.1 in}
where 
\vspace{0.1 in}
\begin{equation}
\fl B\left(\{X_{l}\}\right) = \exp\left[-
\sum_{l=1, l \neq k-1,k,k+1}^N \frac{\left(X_l - X_{l-1}\right)^2}{2
\sigma_l^2}
\right], \;\;\; X_0 \equiv 0,
\end{equation} \vspace{0.1 in}
and
\vspace{0.1 in}
\begin{equation}
\fl B'\left(X_{k-1},X_k,X_{k+1}\right) = \exp\left[- \sum_{l=k-1}^{k+1}\frac{\left(X_{l} - X_{l-1}\right)^2}{2
\sigma_{l}^2} \right].
\end{equation} \vspace{0.1 in}
Note now
that, trivially,
\vspace{0.1 in}
\begin{equation}
\fl \exp\left[ - \frac{\left(X_{k-1} - X_{k-2}\right)^2}{2
\sigma_{k-1}^2} - \frac{\left(X_{k+1} - X_{k}\right)^2}{2
\sigma_{k+1}^2}\right] \leq 1,
\end{equation}
and hence, $P_N$ is bounded from above by
\begin{eqnarray}
\label{gg} \fl P_N \leq  \int_{-L}^{L}
\ldots \int_{-L}^{L} \prod_{l=1, l\neq k-1, k}^{N} \frac{dX_l}{\sqrt{2 \pi} \sigma_l} \,  B\left(\{X_{l}\}\right) \times \nonumber\\
\times \left(\int_{-L}^{L} \frac{dX_k}{\sqrt{2 \pi} \sigma_k} \int_{-L}^{L}
\frac{dX_{k-1}}{\sqrt{2 \pi} \sigma_{k-1}} \exp\left[ - \frac{\left(X_k - X_{k-1}\right)^2}{2
\sigma_k^2}\right]\right)
\end{eqnarray}
Performing integration over $X_k$ and $X_{k-1}$, and differentiating
both sides of the inequality in Eq.(\ref{gg}) with respect to $\sigma_k$, we
get
\vspace{0.1 in}
\begin{eqnarray}
\label{ggg} \fl \frac{\partial P_N}{\partial \sigma_k} \leq - \frac{1}{\pi \sigma_{k-1}} \left(1 - \exp\left[-
\frac{2 L^2}{\sigma_k^2}\right]\right) \int_{-L}^{L}
\ldots \int_{-L}^{L}  \prod_{l=1, l\neq k-1, k}^{N} \frac{dX_l}{\sqrt{2 \pi} \sigma_l} \, B\left(\{X_{l}\}\right) \leq 0
\end{eqnarray} 
This proves the inequality in Eq.(\ref{in}).

\subsection{Bounds: fBm as a Brownian motion with variance $\sigma_N^2$.}

Suppose that we set in Eq.(\ref{f}) all $\sigma_l$ equal to
$\sigma_N$ such that the survival probability in Eq.(\ref{f}) becomes
\vspace{0.1 in}
\begin{eqnarray}
\label{z} \fl P_N' = 
 \int_{-L}^{L} \ldots
\int_{-L}^{L} \prod_{l=1}^{N} \frac{dX_l}{\sqrt{2 \pi} \sigma_N} \exp\left[- \sum_{l=1}^N
\frac{\left(X_l - X_{l-1}\right)^2}{2 \sigma_N^2} \right], \;\;\; X_0 \equiv 0.
\end{eqnarray} \vspace{0.1 in}
In virtue of the fundamental property (ii) of $P_N$, for subdiffusion, i.e., for $H$ such that $0 < H < 1/2$, we have the following \textit{upper} bound:
\vspace{0.1 in}
\begin{equation}
\label{m1}
\fl P_N \leq P_N',
\end{equation} \vspace{0.1 in}
since we have replaced $\sigma_l$, (which is for subdiffusion a \textit{monotonically
decreasing} function of $l$), by its lowest possible value
$\sigma_N$. On contrary, for superdiffusion we have an inverse inequality, i.e., a \textit{lower} bound,
\vspace{0.1 in}
\begin{equation}
\label{m2}
\fl P_N \geq P_N',
\end{equation} \vspace{0.1 in}
since in the superdiffusion case $\sigma_l$ is a \textit{monotonically increasing}
function of $l$ and thus we have replaced it by its highest
possible value.

Now, one notices that $P_N'$ describes the probability $P\left({\rm
max|X_t^{H=1/2}| \leq M}\right)$ that an $N$-step Brownian trajectory, starting at
the origin, does not leave an interval $[-M,M]$,
where $M = L/\sigma_N$. Hence, in virtue of Eq.(\ref{w}), we have
\vspace{0.1 in}
\begin{eqnarray}
\label{ww} \fl P_N' \sim \exp\left[- \frac{\pi^2 \sigma_N^2}{8 L^2}
N\right] &\sim& \exp\left[- \frac{\pi^2}{8\Gamma^2(H+1/2)}
\frac{(\Delta N)^{2 H}}{L^2}\right] = \nonumber\\
&=& \exp\left[- 2 H \frac{\pi^2}{8} \frac{E\left[\left(X_t^H\right)^2\right]}{L^2}
 \right]
\end{eqnarray} 

The result in Eq.(\ref{ww}) defines a \textit{lower} bound on the
survival probability $P_N$ in the superdiffusive case, and
simultaneously, represents an \textit{upper} bound on $P_N$ for
subdiffusion, $0 < H < 1/2$.

\subsection{Bounds: fBm as a Brownian motion in time $t^{2 H}$.}

Consider finally a lower and an upper bounds on $P_N$ which still rely on the fundamental property (ii) of the survival probability $P_N$ but also involve a little bit different type of arguments.

In essence, in this subsection we proceed to show that $P_N$ can be bounded by
\vspace{0.1 in}
\begin{equation}
\fl \int_{-L}^{L} P_{lb}(X,t) dX \leq P_t \leq \int_{-L}^{L} P_{ub}(X,t) dX,
\end{equation} \vspace{0.1 in}
where $P_{lb}(X,t)$ and $P_{ub}(X,t)$ obey:
\vspace{0.1 in}
\begin{equation}
\label{r1}
\fl \frac{\partial P_{lb}(X,t)}{\partial t} = \frac{1}{2 T_{lb}(t)} \frac{\partial^2 P_{lb}(X,t)}{\partial X^2}, \; P_{lb}(X,t=0) = \delta(X), \; T_{lb}(t) = \frac{\Gamma^2(H+1/2)}{t^{2 H - 1}},
\end{equation}
\vspace{0.1 in}
\begin{equation}
\label{r2}
\fl \frac{\partial P_{ub}(X,t)}{\partial t} = \frac{1}{2 T_{ub}(t)} \frac{\partial^2 P_{ub}(X,t)}{\partial X^2}, \; P_{ub}(X,t=0) = \delta(X), \; T_{ub}(t) = \frac{\Gamma^2(H+3/2)}{2 H \, t^{2 H - 1}},
\end{equation}
which have to be solved subject to the boundary condition:
\vspace{0.1 in}
\begin{equation}
\fl P_{lb,ub}(X = \pm L,t) \equiv 0
\end{equation} \vspace{0.1 in}
In other words, we will show that survival probability $P_N$ can be bounded by survival probabilities of Brownian motions evolving in time $t^{2 H}$. 

Note, that equations similar to 
Eqs.(\ref{r1}) or (\ref{r2}) have been 
already proposed in the literature \cite{lung} as some \textit{effective} differential equations describing fractional Brownian motion. However, despite the fact that Eq.(\ref{r1}) reproduces correctly the variance, Eq.(\ref{v}), the Green's function, and, as we will see, the time-dependence of the survival probability, it is not \textit{exact} and can not reproduce correctly the correlations in the fBm process, Eq.(\ref{e}).

Now, we turn back to our result in Eq.(\ref{f}) and notice that 
 in the general case $0 < H \leq 1$, i.e. for
superdiffusion, diffusion and subdiffusion regimes, and for any $l \geq 1$,
\vspace{0.1 in}
\begin{equation}
\label{zz}
\fl \sigma^2_l \leq \Sigma^2(l) = \frac{\Delta^{2 H}}{2 H\Gamma^2(H+1/2)} \, g(l),
\end{equation} \vspace{0.1 in}
where
\vspace{0.1 in}
\begin{equation}
\label{g}
\fl g(l) = l^{2 H} - (l-1)^{2 H}.
\end{equation} \vspace{0.1 in}
In fact, the inequality in Eq.(\ref{zz}) with $g(l)$ defined by Eq.(\ref{g}) 
appears to be a very good approximation for $\sigma_l$: while $\sigma_l$ and $\Sigma(l)$ show the same behavior as functions of $l$ for sufficiently large $l$, $\sigma_{\infty}/\Sigma(\infty) = 1$, they differ only by a few per cent also for moderate values of $l$.

Now, in virtue of (ii), we have the following \textit{lower} bound
on the survival probability $P_N$:
\vspace{0.1 in}
\begin{eqnarray}
\label{l1}
\fl P_N &\geq&  \int_{-L}^{L} \ldots \int_{-L}^{L}
\prod_{l=1}^{N} \frac{dX_l}{\sqrt{2 \pi} \Sigma(l)} \exp\left[- \sum_{l=1}^N \frac{\left(X_l -
X_{l-1}\right)^2}{2 \Sigma^2(l)} \right] = \nonumber\\
\fl &=& \int_{-M}^{M} \ldots \int_{-M}^{M}
\prod_{l=1}^{N} \frac{dZ_l}{\sqrt{2 \pi (l^{2 H} - (l - 1)^{2 H})}} \exp\left[- \sum_{l=1}^N \frac{\left(Z_l -
Z_{l-1}\right)^2}{2 (l^{2 H} - (l - 1)^{2 H})} \right],
\end{eqnarray} \vspace{0.1 in}
where $ Z_0 \equiv 0$ and $M = \sqrt{2 H} \, \Gamma(H + 1/2) L/\Delta^H$.

One notices next that the expression in the second line in Eq.(\ref{l1}) defines the probability that an $N$-step trajectory of Brownian motion, starting at the origin and evolving in \textit{time} $T = (\Delta N)^{2 H}$ does not escape from the interval $[-M,M]$. Hence, in virtue of Eq.(\ref{w}), the survival probability $P_N$ is bounded from below by
\vspace{0.1 in}
\begin{eqnarray}
\label{l3}
\fl P_N \geq \exp\left[ - \frac{\pi^2}{16 H \Gamma^2(H+1/2)}
\frac{(\Delta N)^{2 H}}{L^2} \right] 
= \exp\left[- \frac{\pi^2}{8} \frac{E\left[\left(X_t^H\right)^2\right]}{L^2}
 \right]
\end{eqnarray}
We emphasize that this bound holds for any $H \in ]0,1]$ and thus applies to both subdiffusion and superdiffusion regimes. Note also that it coincides with the results in Eqs.(\ref{lll}) and (\ref{psi}).

Next, note that for any $l \geq 1$ and any $H \in ]0,1]$, we have
\vspace{0.1 in}
\begin{equation}
\sigma^2_l \geq \tilde{\Sigma}^2(l) = \frac{\Delta^{2 H}}{\Gamma^2(H+3/2)} \, g(l),
\end{equation}
where $g(l)$ is defined by Eq.(\ref{g}).

Hence, in virtue of (ii), we have the following \textit{upper} bound
on the survival probability $P_N$:
\vspace{0.1 in}
\begin{eqnarray}
\label{l2}
\fl P_N &\leq&  \int_{-L}^{L} \ldots \int_{-L}^{L}
\prod_{l=1}^{N} \frac{dX_l}{\sqrt{2 \pi} \tilde{\Sigma}(l)} \exp\left[- \sum_{l=1}^N \frac{\left(X_l -
X_{l-1}\right)^2}{2 \tilde{\Sigma}^2(l)} \right] = \nonumber\\
\fl &=& \int_{-M}^{M} \ldots \int_{-M}^{M}
\prod_{l=1}^{N} \frac{dZ_l}{\sqrt{2 \pi (l^{2 H} - (l-1)^{2 H})}} \exp\left[- \sum_{l=1}^N \frac{\left(Z_l -
Z_{l-1}\right)^2}{2 (l^{2 H} - (l - 1)^{2 H})} \right],
\end{eqnarray} \vspace{0.1 in}
where $Z_0 \equiv 0$ and $M = \Gamma(H + 3/2) L/\Delta^H$. Consequently, 
\begin{eqnarray}
\label{l4}
\fl P_N \leq \exp\left[ - \frac{\pi^2}{8  \Gamma^2(H+3/2)}
\frac{(\Delta N)^{2 H}}{L^2} \right] 
= \exp\left[- \frac{2 H}{\left(H + 1/2\right)^2} \frac{\pi^2}{8} \frac{E\left[\left(X_t^H\right)^2\right]}{L^2}
 \right]
\end{eqnarray} 
This bound also holds for any $H \in ]0,1]$, i.e. for both superdiffusion and subdiffusion.

Note now that bounds in Eqs.(\ref{l3}) and (\ref{l4}) appear to be sharper than those defined by Eqs.(\ref{m1}),(\ref{m2}) and (\ref{ww}). Indeed, for superdiffusion the bound in Eq.(\ref{l3}) is higher than the one defined by Eqs.(\ref{m2}) and (\ref{ww}), since here $2 H > 1$. For subdiffusion, the upper bound in Eqs.(\ref{m1}) and (\ref{ww}) is also worse, i.e. higher, than the one defined by Eq.(\ref{l4}) since $2 H < 2H/(H+1/2)^2$ for $H < 1/2$.

Therefore, the main result of the present paper can be represented as the 
following double-sided inequality on $P_N$:
\vspace{0.1 in}
\begin{equation}
\frac{2 H}{(H + 1/2)^2} \leq - \left(\frac{8}{\pi^2} \frac{L^2}{E\left[\left(X_t^H\right)^2\right]}\right) \ln\left(P_N\right) \leq 1,
\end{equation}
which holds for any $H \in ]0,1]$. Note that the bounds on the right-hand and on the left-hand-side coincide, as they should, for $H = 1/2$.

\section{Conclusions}

To conclude, we have studied  the long-time asymptotical behavior of the 
probability $P_t$ 
that the Riemann-Liouville
fractional Brownian motion with Hurst index $H$ 
does not escape from a fixed interval $[-L,L]$ up to time $t$. We have shown that $P_t$
 obeys $\ln(P_t) \sim - t^{2 H}/L^2$. This result is valid for 
any $H \in ]0,1]$, 
for both subdiffusion ($0 < H < 1/2$) and superdiffusion ($1/2 < H \leq 1$) regimes, and consequently,  
the decay  may be slower than exponential (subdiffusion) or faster than exponential (superdiffusion). 

This decay law has been obtained by a) showing that for $1/4 < H \leq 1$ the survival probability $P_t$ of the fBm in presence of two fixed adsorbing boundaries is determined by
 the probability that a Brownian motion does escape from the interval with moving boundaries and b) by  elaborating upper and lower bounds on $P_t$ which show the same time dependence. 
These bounds stem from some fundamental property of the survival probability, Eq.(\ref{in}), and controllable approximation of the fractional Brownian motion by standard Brownian motion evolving in time $T = t^{2 H}$. 

The obtained result for the survival  probability decay implies, in particular, that the 
survival probability $S_t$ of particles undergoing fBm in
one-dimensional systems with randomly placed traps obeys $\ln(S_t) \sim - n^{2/3} t^{2H/3}$, where $n$ is the mean density of traps. This expression generalizes famous Balagurov and Vaks \cite{bal} and Donsker and Varadhan \cite{don}
result over the case of anomalous diffusion described as fractional 
Brownian motion. For $H = 1/4$, we find the $\ln(S_t) \sim - t^{1/6}$ law, 
which was previously obtained in Ref.\cite{nech}.

Finally, we note that the analysis presented in this paper can be
straightforwardly generalized to fBm
taking place in higher-dimensional spaces, 
the case of Weyl fractional
Brownian motion, as well as for evaluation of bounds on 
the distribution function of the
 range of fBm, and on the survival probability of the fBm in
presence of one-sided or two-sided moving boundaries.

\section{Acknowledgments}

The author acknowledges helpfull discussions with P.Krapivsky,
A.Grosberg, S.Nechaev and M.Kardar.

\end{document}